\documentclass[aps,prl,amssymb,reprint,twocolumn,superscriptaddressgroupaddress]{revtex4-1}
\usepackage{amsmath}
\usepackage{hyperref}
\usepackage[english]{babel}
\usepackage[utf8]{inputenc}
\usepackage{float}
\usepackage{graphicx}
\usepackage{braket}
\usepackage{xcolor}
\usepackage{lipsum}

\begin{document}
	\title{Moir\'{e} versus Mott: Incommensuration and Interaction\\
		 in One-Dimensional Bichromatic Lattices }
	\author{DinhDuy Vu}
	\author{S. Das Sarma}
	\affiliation{Condensed Matter Theory Center and Joint Quantum Institute, Department of Physics, University of Maryland, College Park, Maryland 20742, USA}

    \begin{abstract}
    Inspired by the rich physics of twisted 2D bilayer moir\'{e} systems, we study Coulomb interacting systems subjected to two overlapping finite 1D lattice potentials of unequal periods through exact numerical diagonalization.  Unmatching underlying lattice periods lead to a 1D bichromatic `moir\'{e}' superlattice with a large unit cell and consequently a strongly flattened band, exponentially enhancing the effective dimensionless electron-electron interaction strength and manifesting clear signatures of enhanced Mott gaps at discrete fillings.  An important non-perturbative finding is a remarkable  fine-tuning effect of the precise lattice commensuration, where slight variations in the relative lattice periods may lead to a suppression of the correlated insulating phase, in qualitative agreement with the observed fragility of the correlated insulating phase in twisted bilayer graphene.  Our predictions, which should be directly verifiable in bichromatic optical lattices, establish that the competition between interaction and incommensuration is a key element of the physics of moir\'{e} superlattices.
   \end{abstract}
   
   \maketitle
   
   \textit{Introduction -} Motivated by the intriguing and interesting recent experimental and theoretical studies observing correlated insulating phases in 2D moir\'{e} systems \cite{Cao2018, Cao2018a, Yankowitz2019, Lu2019,Regan2020,Jin}, we ask a simple conceptual question: Is there a one-dimensional analog for interacting moir\'{e} superlattices where correlated insulating phases manifest strongly in superlattices at fractional band fillings, but not in the corresponding original lattice? The advantage of asking this question in a  1D system is that the problem can be studied using exact numerical  diagonalizations, something completely out of question in 2D moir\'{e} systems because of the exponentially large Hilbert space.  In the current work, we provide detailed numerical results answering this question by focusing on electrons in a bichromatic 1D lattice with superposed periodic potentials with equal amplitudes, but differing lattice sizes. Here, the lattice size ratio defining the moir\'{e} superlattice through its rational fractional representation.  Our work establishes definitively that indeed correlated insulating phases emerge generically in such moir\'{e} superlattices, and in addition, we also show that the emergent correlated insulating phase is fragile, and may disappear under slight variations in the lattice period ratio because of a subtle competition between incommensuration and interaction. Our results, in addition to being of possible relevance to the observed correlated insulating phase in 2D moir\'{e} systems, should also be directly observable in 1D bichromatic atomic optical lattices \cite{Schreiber2015,Li2017,Luschen2018,Kohlert2019}
   
   \textit{Model -} The model we study is a bichromatic 1D system with two overlapping periodic cosine potentials of periods $a_1$ (which is taken to be the primary lattice) and $a_2$, so the single-particle lattice potential $V(x)$ is given by $V(x)=V_0\left[ \cos(2\pi x/a_1) +\cos(2\pi x/a_2) \right]$. Here $V_0<0$ is a constant defining the lattice potential strength, and $a_2/a_1=m$ defines the moir\'{e} ratio of the combined potential. The Coulomb interaction between the electrons at the locations $x_i$ and $x_j$ is given by $U(x,x') = e^2/|x-x'|$. We take $a_1/a_B=r_s$ as the dimensional lattice length, where $a_B$ is the Bohr radius (all length is measured in units of Bohr radius and all energies in units of Hartree $E_h=e^2/a_B$). The system Hamiltonian is now defined by $V + U$, which we solve numerically by the Density Matrix Renormalization Group method \cite{Stoudenmire2012,itensor} for a system with $N_e$ particles on a total system size of $L=Na_1$ (i.e. $N$ primary lattice sites) with open boundary conditions \cite{Appendix}. 
   
   The ratio $m=a_2/a_1$ is the key moir\'{e} superlattice parameter in our model, serving a role similar to the twist-angle $\theta$ in twisted 2D moir\'{e} heterostructures. While the unit cell size in the original lattice is $a_1$, in the moir\'{e} superlattice the unit cell size is increased to $A=Ma_1 > a_1$, where $M$ is the lowest possible integer numerator in the rational fraction representation of $m$.  For example, with $a_2=1.4a_1$, $1.5a_1$, $1.6a_1$, three examples used in this work, $M=7$, 3, 8 respectively, whereas for the original lattice $(a_1=a_2)$, $m=M=1$.  The enhanced value of $A$ compared with $a_1$ leads to flat bands in the 1D superlattice very similar to what happens in 2D twisted systems, e.g. twisted bilayer graphene (tBLG) \cite{Cao2018, Cao2018a, Yankowitz2019, Lu2019} or transition metal dichalcogenide (TMD) heterostructures \cite{Regan2020,Jin}.
   
   \textit{Results and Discussions -} In the rest of this paper, we now present and discuss our results for the 1D superlattice moir\'{e} states for $m=1.4$, 1.5, 1.6 cases, comparing with the original non-moir\'{e} situation of $m=1$.  Unless otherwise stated we choose $V_0=-3.5~E_h$, $L=63a_1$, $N_e=1-10$.  We define $U_c=e^2/A$ as the Coulomb coupling strength and $t=(2A^2)^{-1}\partial^2 E(k)/\partial k^2$ at $k=0$ as the kinetic energy strength, thus providing $U_c/t$ as the dimensionless interaction strength. Note that $U_c/t$ depends on $r_s=a_1/a_B$, and $U_c/t$ increases  exponentially, for fixed $r_s$, with increasing $A$ since $t$ is suppressed exponentially, but $U_c$ only as $1/A$, in the superlattice.  We provide the calculated $U_c/t$ values for each interacting situation we study in our results along with the values of $m$ and $M$. Our results are mostly for $r_s=1$ unless otherwise stated explicitly.  Note that $r_s=1$ indicates a situation (for the original lattice, $a_1=a_B$) which is of intermediate coupling strength since the strong coupling regime is defined by $r_s\gg1$.  We note that the values of $r_s$ and $U_c/t$ are inferred quantities provided as a context and are not used anywhere in our calculations, which are all exact using the superlattice single-particle cosine potentials $V(x)$ and the inter-particle $1/x$ Coulomb interaction.
   
\begin{figure}
   	\includegraphics[scale=0.5]{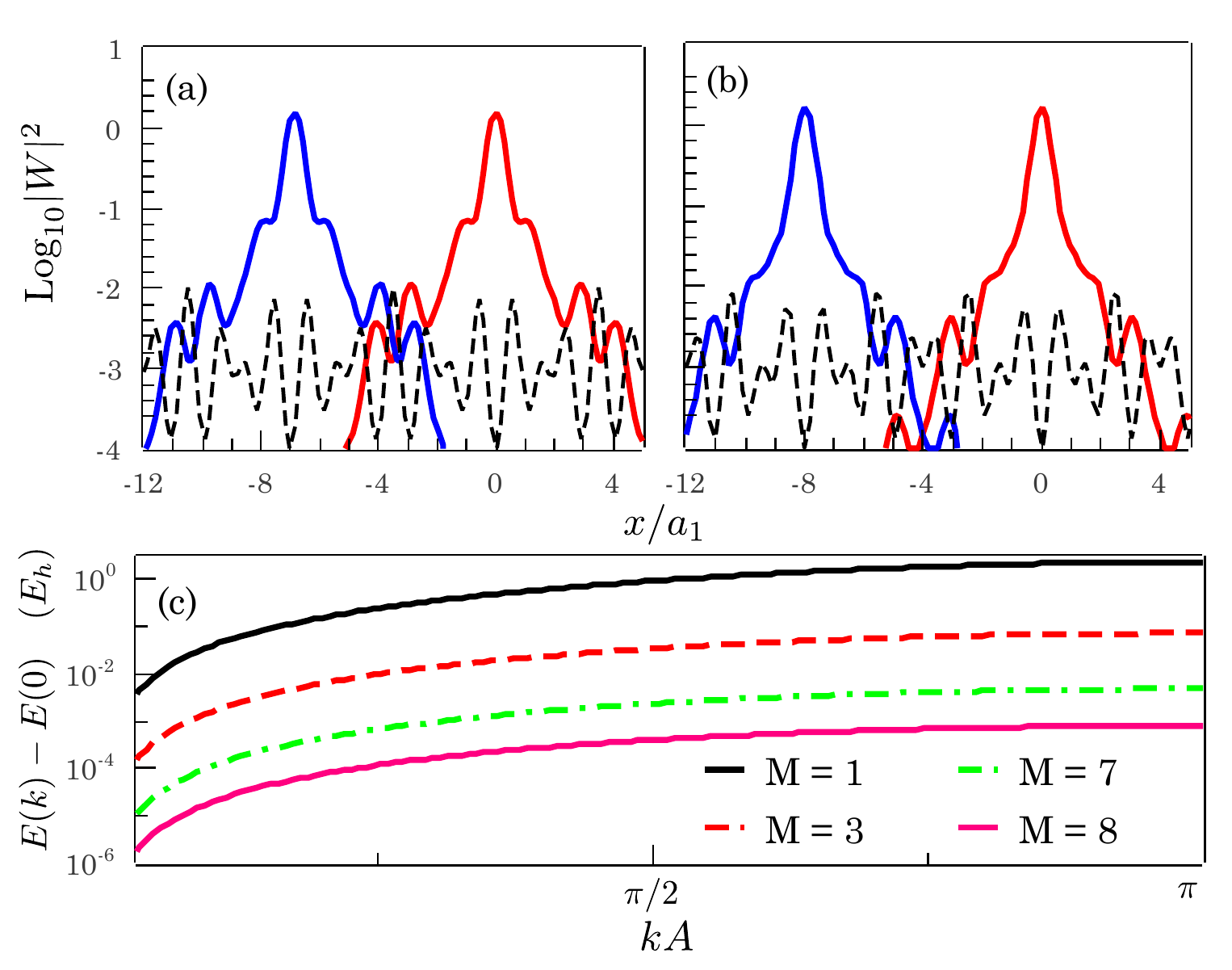}
   	\caption{Squared amplitude $|W|^2$ of two adjacent Wannier wavefunctions for (a) $M = 7$ ($m=1.4$) and (b) $M = 8$ ($m=1.6$). The dashed black lines are the total background potential including the primary and the incommensurate lattice fields. (c) The lowest band dispersion as the function of the moir\'{e} momentum. \label{fig1}}
   \end{figure}

   In Fig.~\ref{fig1}, we present the exactly calculated noninteracting band structure of the infinite 1D superlattice for different values of $m$, clearly showing an exponential flattening of the bands with increasing $M$. Some representative non-interacting Wannier functions are also shown to emphasize the almost `localized' nature of the noninteracting superlattice band states for $m=1.4$ ($M=7$) and 1.6 ($M=8$), similar to the 2D twistronic materials \cite{Angeli2020,Carr2020,Naik2020}.  Thus, impressive band flattening is already achieved for a modest incommensuration of $m=1.4$ or 1.6.
   
\begin{figure}
   	 \includegraphics[scale=0.6]{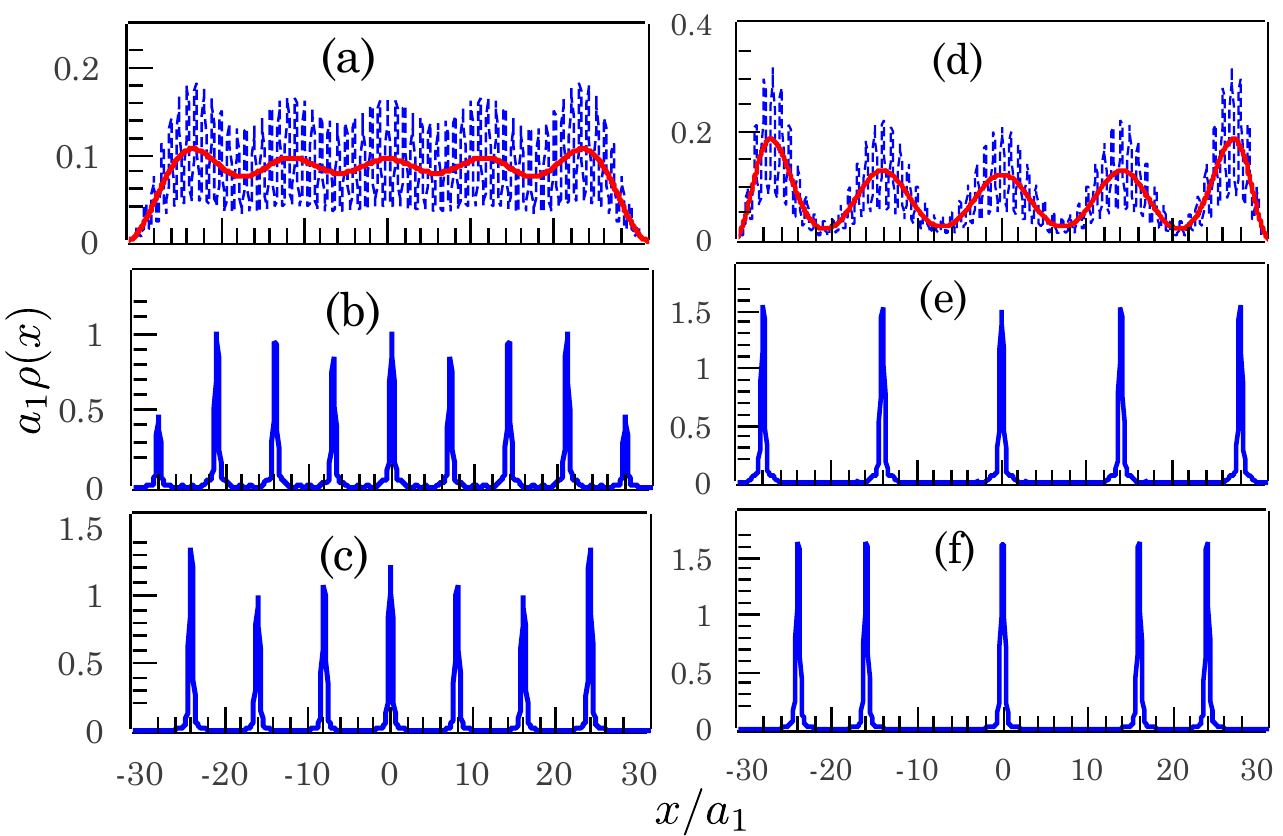}
   	 \caption{Spatial density distribution for 5 non-interacting
   	 	electrons in a lattice field with (a) $M = 1$, (b) $M = 7$, (c) $M = 8$. (d)-(f) interacting counterpart of (a), (b) and (c), respectively. The red lines are smoothed functions of the actual density distributions (dashed lines) \label{fig2}}
   \end{figure}

   In Fig.~\ref{fig2}, we show the calculated density distributions in the interacting moir\'{e} cases ($M=7$, 8) compared with the corresponding noninteracting situations for $N_e=5$. It is obvious that while all sites have finite occupancies in the noninteracting situation (there are $63/M$ sites in the moir\'{e} system) indicating a standard metallic band, the corresponding interacting situation is strongly localized with only 5 density peaks corresponding to the 5 particles in the system.  For comparison, we also show the results for the original lattice ($M=1$), where the localization is substantially weaker than for $M=7$, 8 showing the strong enhancement of correlation effects in the interacting superlattice system. The corresponding dimensionless interaction strengths, as obtained from our calculated band widths,  increase from $U_c/t=2.5$ for $M=1$ to 123.6 ($M=7$) and 618.8 ($M=8$).  Clearly, the moir\'{e} superlattice enables the manifestation of the localized insulating correlated states in a generic manner.
   
\begin{figure}
   	\includegraphics[scale=0.5]{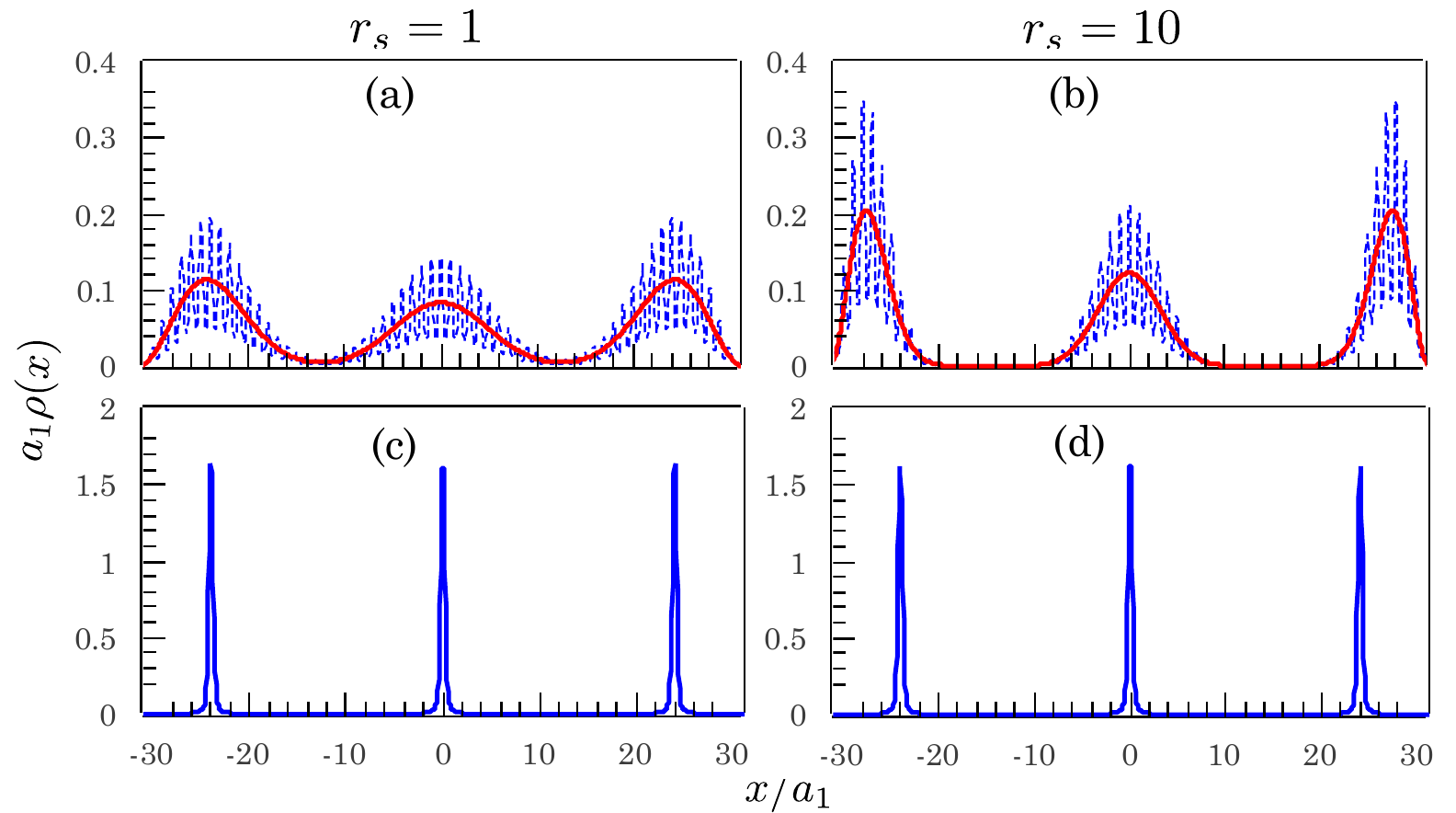}
   	\caption{Density-induced localization in 3-electron systems from $r_s = 1$ [(a) and (c)] to $r_s = 10$ [(b) and (d)]. (a),(b) Original lattice potential $M = 1$ with $U_c/t = 2.5$ and 25 for $r_s =1$ and 10 respectively. The red lines are smoothed functions of the actual density distributions (dashed lines). (c),(d) Incommensurate lattice potential $M = 8$ with $U_c/t = 618.8$ and 6188 for the two values of $r_s$. \label{fig3}}
   \end{figure}

   In Fig.~\ref{fig3}, we show that the correlation physics in the superlattice is much more akin to Mott localization \cite{Mott1949} (driven by $U_c/t$ within narrow band physics) than electron gas Wigner crystallization \cite{Wigner1934a} (i.e. driven by increasing $r_s$, the dimensionless density of the system).  Changing $r_s$ from 1 to 10 does not change the density profile at all for the interacting superlattice ($M=8$), but it affects the density profile in the original lattice ($M=1$) as increasing $r_s$ enhances $U_c/t$ from a smaller value to a larger one (by contrast, the superlattice already has exponentially high $U_c/t$). We note that beside the nomenclature difference, our Mott insulator and the `Wigner crystal' (correlated insulating phase at less than one particle per site) terminology \cite{Wu2007,Regan2020,Jin} share the same origin of narrow-band electrons interacting via a long-range interaction. We are using `Mott' to generically signify any interaction-driven fractional filling insulating ground state in a lattice system.
   
\begin{figure}
   	\includegraphics[width=0.43\textwidth]{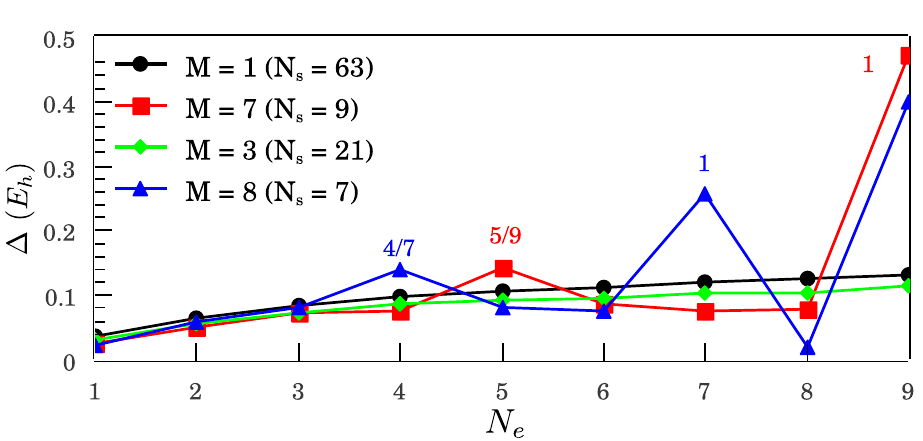}
   	\caption{Charge gap as a function of the electron number $N_e$ for
   		lattice potentials with various degrees of incommesuration. 
   		Some filling factors $N_e/N_s$ are indicated with $N_s$ being the number 
   		of moir\'{e} sites. \label{fig4}}
   \end{figure}

   In Fig.\ref{fig4}, we show the calculated charge gap $\Delta(N_e)=E(N_e+1)+E(N_e-1)-2E(N_e)$ for different $M$ values plotted as a function of the electron number (or equivalently, band filling), clearly showing that a correlated insulating gap emerges in the superlattice (for $M=7, 8$) where the corresponding non-superlattice case ($M=1,$ 3) does not manifest any pronounced insulating behavior.  The overall smooth increase of the gap with $N_e$ is the so-called collective Coulomb blockade behavior where the gap increases smoothly in any finite system as the number of electrons increases in it because of the standard Coulomb repulsion physics \cite{Stafford1994,Hensgens2017}.  The real correlated gaps of interest are the peaks (at half-filling for $M=7$, 8) above the smooth background, and one should subtract out the smooth background to get the true correlated gap.  The moir\'{e} system manifests a large correlated gap at the commensurate filling of 1/2 (and in fact, there are smaller gaps at other commensurate fillings $1/n$ \cite{Appendix}, but severe finite size effects overwhelm those gaps in the results of Fig.~\ref{fig4}).  Note that while this correlated gap arises for $m=1.4$ ($M=7$) and $m=1.6$ ($M=8$), it is manifestly absent at the intermediate $m$-value of 1.5 ($M=3$).  This is a clear indication of the interplay between moir\'{e} superlattice and Coulomb interaction -- the $M=3$ ($m=1.5$) case has very weak band flattening effect, and hence very weak correlation gap. 
   
   The manifestation of correlated gap is nonmonotonic in the value of $m$, it is enhanced when $m$ approaches an irrational such as $\sqrt{2} \sim 1.414\dots$ and Golden mean $\sim 1.618\dots$ for $m=1.4$ and 1.6 respectively because the rational approximation to such irrationals involves a large value of $M$, concomitantly flattening the single-particle bands, and hence enhancing the dimensionless coupling strength $U_c/t$ very strongly.  This does not happen at the simple rational fraction $m=1.5=3/2$ (i.e. $M=3$) where the band flattening and hence the insulating gap effects are weak.  We believe this same phenomenon is reflected in 2D moir\'{e} superlattices where the correlated insulating gaps are fragile because the twist angle $\theta$ varies somewhat over the sample size, and sometimes it simply produces rational bands without enhancing the correlation effects.
   
\begin{figure}
   	\includegraphics[scale=0.4]{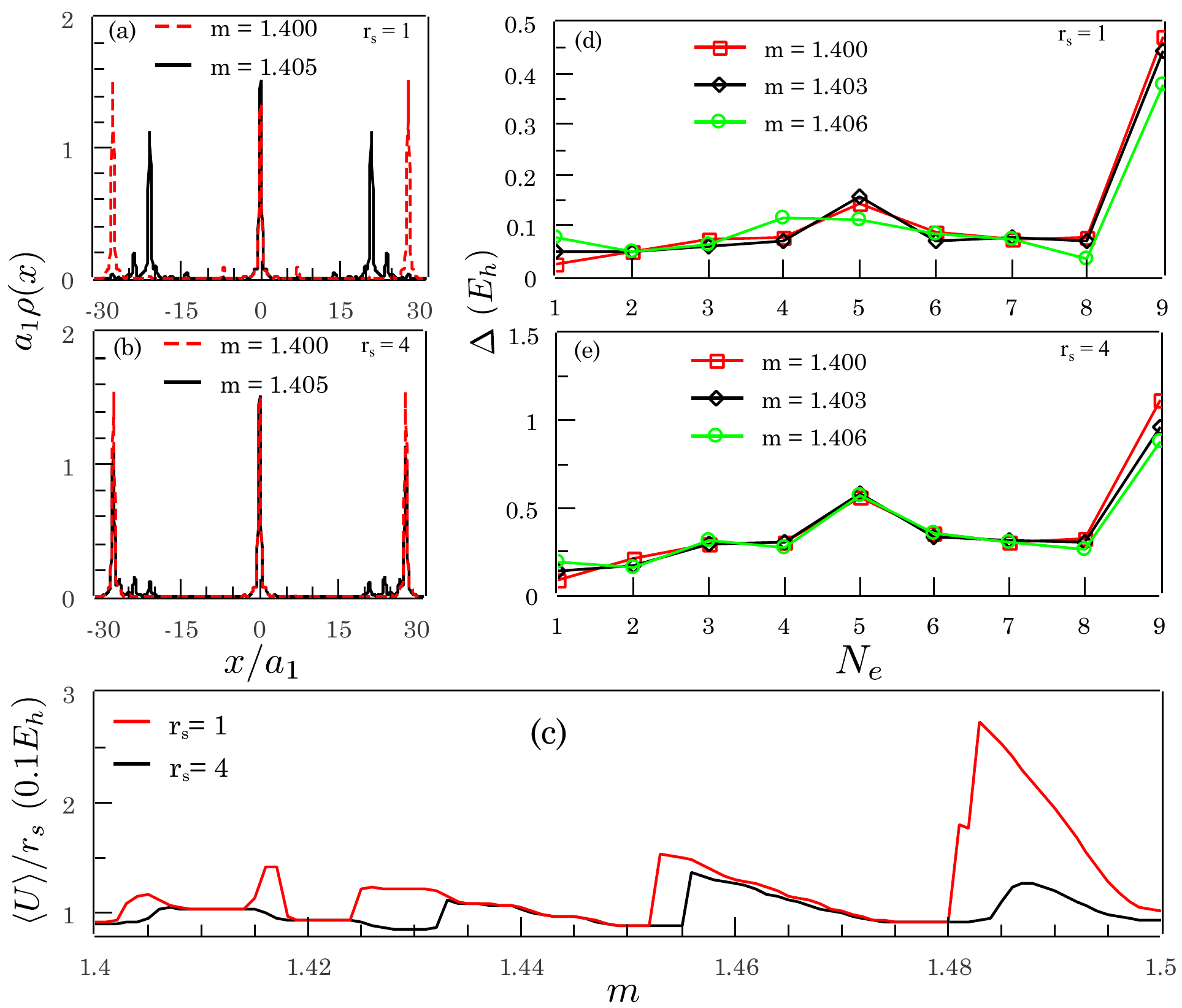}
   	\caption{(a),(b) The spatial configuration of 3 electrons on the periodic ($m=1.400$) and disordered ($m=1.405$) superlattice arrays for $r_s=1$ and $r_s=4$. (c) Average interaction energy of the 3-electron system normalized by $r_s$, as a function of the lattice commensuration $m$, where the peaks indicate the absence of correlated insulator. (d),(e) Charge gap as a function of the electron number with the half-filling peak  disappearing for $r_s=1$ but stable for $r_s=4$.\label{fig6}}
   \end{figure}

   In this context, we discuss a subtle competition, shown in Fig.~\ref{fig6},  between quasi-periodic incommensuration and interaction, which reflects the delicate fragility of the correlated insulator phase. This physics arises from the incommensurate system with a very large $M$ potentially having a unit cell size $A$ exceeding the system size. Such phenomena can be demonstrated by slightly modifying $m$ by less than 1\% away from $1.4$ while keeping everything else exactly the same. With such fine tuning, e.g from $m=1.4$ to 1.405, $M$ increases from 7 to 200! Since this period is much larger than the system size, the fine tuning of $\delta m=0.005$ does not create a new superlattice pattern but introduces a disordered variation among the original moir\'{e} sites ($A=7a_1$). In general, a complex rational or irrational moir\'{e} ratio induces a quasi-random potential \cite{Appendix} that may replace the Mott phase with a disorder-induced Anderson localization \cite{DasSarma1988,Li2017,Luschen2018}. In Figs.~\ref{fig6}(a), by tuning $m$ from 1.4 to 1.405, the Mott spatial configuration is destroyed as the modified configuration obviously does not minimize the interaction energy. The way to circumvent this disorder-induced physics and revive strong correlation effects is to increase the basic interaction strength $r_s$, as shown in Figs.~\ref{fig6}(b), where for $r_s=4$, the same tuning of $m$ has no effect on the correlated insulating phase. As the Mott insulator minimizes the interaction energy, we show the average interaction energy $\braket{U}$ scaled by $r_s$ in Fig.~\ref{fig6}(c) with continuously varied $m$. Starting with $a_2=1.4a_1$, where the correlated insulator clearly exists,  and going to $a_2=1.5a_1$, the rational approximation of $m$ varies making $M$ sometimes very large and sometimes rather small, consequently almost randomly suppressing [the peaks in Fig.~\ref{fig6}(c)] and enhancing the correlation effects. For larger $r_s$ ($r_s=4$), this incommensuration-induced fragility is less prominent because interaction now overcomes incommensuration. We note that the quasi-periodicity strongly affects low-filling states, thus making fractional-filling gaps more fragile than the integer-filling counterpart. As shown in Fig.~\ref{fig4}(d) the calculated charge gaps for $r_s=1$ and different values of fine-tuned $m$, the half-filling peak is washed out as $m$ is tuned while the full-filling peak is relatively robust. The fractional-filling correlated insulator, however, can be stabilized by a stronger basic interaction strength, e.g. for $r_s=4$ as in Fig.~\ref{fig4}(e). We believe this disorder caused by the finite size is related to the observed fragility of the correlated insulated phase, especially at the filling of less than one particle per moir\'{e} site.
   
\begin{figure}
	\includegraphics[scale=0.55]{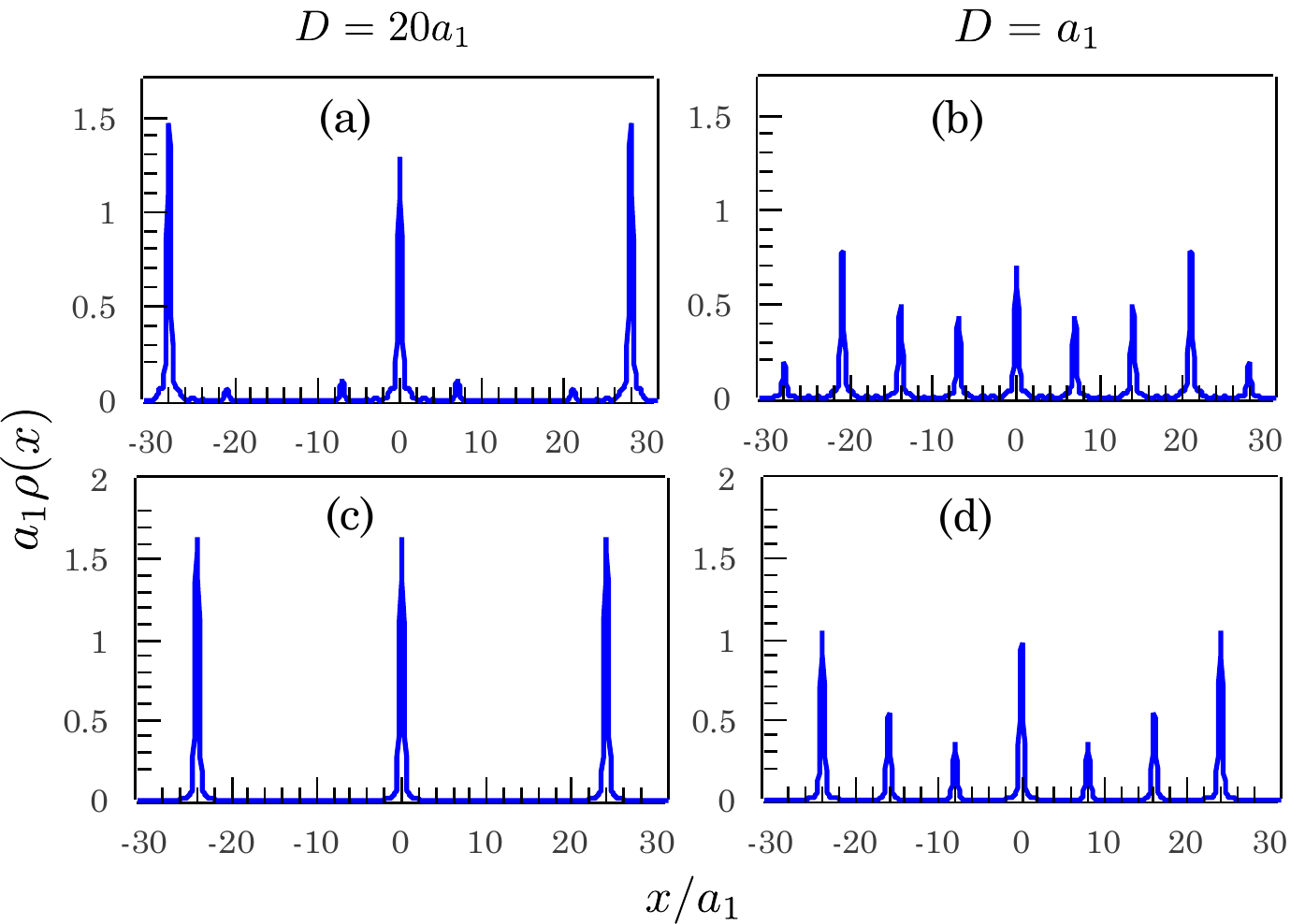}
	\caption{Screening-induced delocalization for 3-electron systems from the screening length $D = 20a_1$ [(a) and (c)] to $D = a_1$ [(b) and (d)]. (a),(b) Lattice potential with $M = 7, U_c/t = 82.8$ and 1.2. (c),(d) Lattice potential with $M = 8, U_c/t = 389.0$ and 4.8. \label{fig5}}
\end{figure}

   One direct way to control the Coulomb coupling is by using an external gate to screen the interaction as already demonstrated in tBLG \cite{Stepanov2019,Saito2020,Liu2020}.  In Fig.~\ref{fig5}, we show the effects of external screening  by a gate placed at a distance of $D$ so that the screened Coulomb interaction between two electrons separated by a distance $x$ gets suppressed to $x^{-3}$ for $x \gg D$ instead of the usual $1/x$ dependence.  Since the period $A$ ($Ma_1)\gg a_1$ controls the moir\'{e} length scale, the gating effect is strong in the moir\'{e} system for $D<A$.  This can be clearly seen in Fig.~\ref{fig5} where we show the calculated density distribution for $M=7$ and 8, which, for small $D=a_1$, becomes essentially metallic with all sites are partially occupied as in the weakly interacting system.

   \textit{Conclusion -} We have carried out finite-size exact calculations for interacting 1D bichromatic lattices, showing that correlated insulating states manifest generically  when the moir\'{e} supelattice has a large unit cell (e.g. $m=1.4$ and 1.6) compared with the original lattice, but not when the two lattices are almost commensurate (e.g. $m=1.5$). This phenomena should apply directly to 1D bichromatic lattices. Beyond the ground state, disordered dynamic phenomena such as quantum chaos and many-body localization \cite{Tian2005,Taniguchi1993,Schreiber2015,Li2017,Luschen2018,Kohlert2019} can also be studied as our model can tune the disorder (through the secondary lattice potential) and interaction (through screening). Our work is also relevant to recent experiments \cite{Li2020} observing 1D anisotropic band flattening in 2D graphene due to an applied superlattice potential, except we estimate, using Ref.~\cite{Brey2009}, that the enhancement of the effective $U_c/t$ here to be algebraic (and not exponential as in our pure 1D moir\'{e} case) and of the $\mathcal{O}$(100) for an induced superlattice period of 100~nm. 
   
   In connection with 2D moir\'e patterns, we cannot comment on the magnetic properties of tBLG and related 2D moir\'{e} systems based on our work.  We also cannot discuss topological properties of 2D moir\'{e} systems based on our work since we have not incorporated any topology in our model. The finite size and dimension of the problem also prevent quantitative analogy with 2D twisted bilayer systems. However, we speculate that our findings qualitatively explain several features of the correlated insulating phase in 2D moir\'{e} systems including gate-induced suppression of the insulator and the generic fragility of the correlated insulating phase. These features arise specifically from the competition between flatband incommensurate moire physics and Coulomb interactions. Indeed, the manifestation of correlated physics in 2D is qualitatively similar to 1D with fractional-filling charge gaps appear generically in narrow band models with long-range interaction \cite{Wu2007}. In addition, recent experiments in TMD reported $1/3, 2/3$ \cite{Regan2020} and $1/2,1/4,2/5,3/5-$filling \cite{Jin} (per moir\'{e} site) correlated insulating phases. These fractional-filling gaps share the same physics with the half-filling gap observed in our model, which suggest a possible 1/8-filling (one electron per two moir\'e sites) in tBLG. However, we expect this insulating phase to be fragile as our 1D simulation establishes that a slight tuning of the moir\'{e} ratio may drastically suppress the insulating phase, especially at fractional fillings, because such tuning may drive the moir\'{e} system from being a periodic superlattice to a disordered system. In fact, we believe that the observed fragility of the tBLG correlated insulating phase may already be reflecting this `moir\'{e} versus Mott' competition described in our work. We also mention that our theory would apply to a general flatband incommensurate interacting situation with more than two overlapping potentials.  Finally, we note that future generalizations of our work, involving studying the whole spectrum (and not just the ground state as done in the current work), could connect with the physics of many body localization and quantum chaos.
  
   \acknowledgements{\textit{Acknowlegement - }
   	  This work is supported by the Laboratory for Physical Sciences.}
   
   	 \bibliographystyle{apsrev4-1}
   	 \bibliography{Moire_paper}

    \appendix
    \widetext  	 
    \section*{Supplemental Material for ``Moir\'{e} versus Mott: Incommensuration and Interaction in One-Dimensional Bichromatic Lattices"}

\section{Numerical method}

\begin{figure}[h!]
	\centering
	\begin{minipage}{0.02\textwidth}
		$E_h$ \vspace{1.2in}
	\end{minipage}
	\begin{minipage}{0.40\textwidth}
		\centering
		\includegraphics[width=\textwidth]{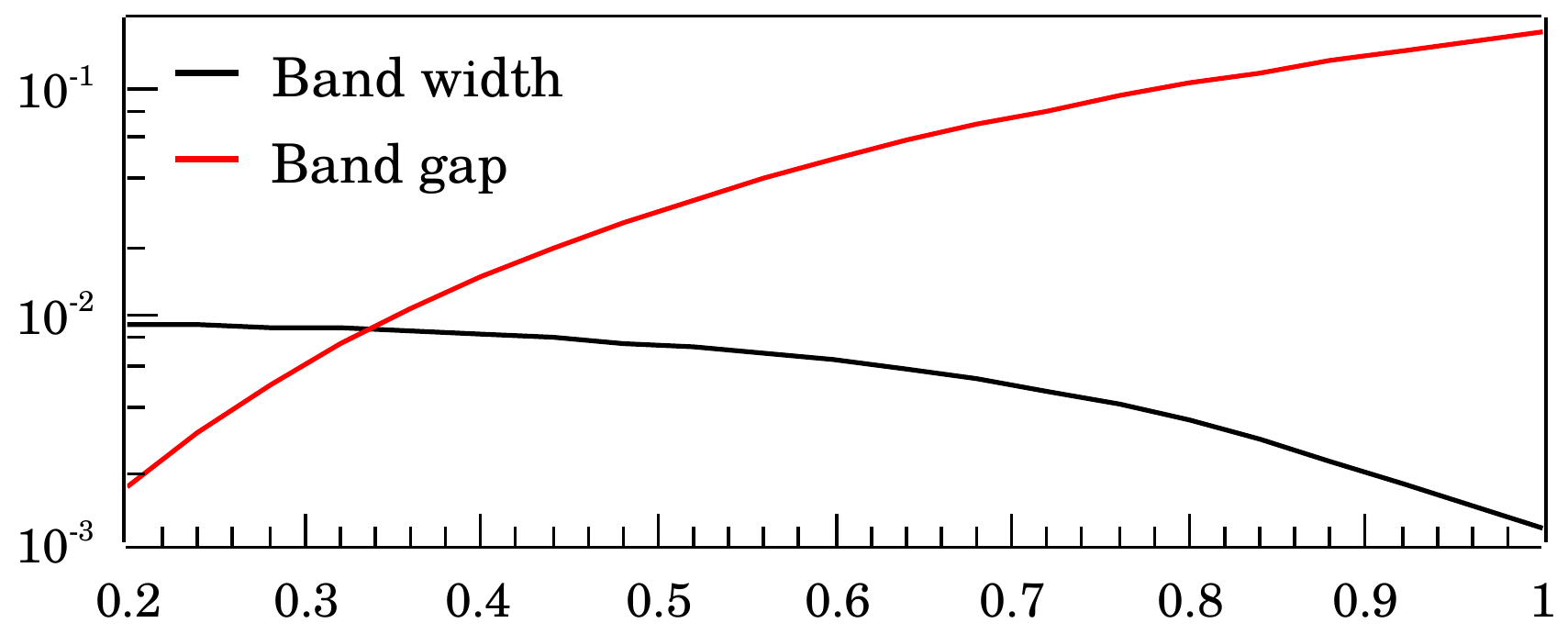}
		
		$V_2/V_1$  	 		
	\end{minipage}
	\caption{Bandwidth and bandgap for the moir\'{e} pattern of $M=7$.\label{fig:chooseV2}}
\end{figure}

We employ the 1D spinless continuum Hamiltonian
\begin{equation}
	H = \int dx \psi^\dagger(x)\left[\frac{-\partial^2}{2\partial x^2}+V_1 \cos\left( \frac{2\pi x}{a_1} \right) + V_2\cos\left( \frac{2\pi x}{a_2}\right) \right]\psi(x) +\frac{1}{2}\int U(x-x')n(x)n(x')dx dx',
\end{equation}
where the open boundary conditions are applied $\psi(-L/2)=\psi(L/2)=0$. The interaction is $U(x)=1/\sqrt{x^2+d^2}$ for the bare Coulomb interaction and $U(x)=1/\sqrt{x^2+d^2} - 1/\sqrt{x^2+D^2}$ with $D$ being the screening length for the screened Coulomb interaction. The continuum Hamiltonian is simulated on a discretized grid with spacing $\Delta$, the discretized Hamiltonian is
\begin{equation}
	\begin{split}
		H = \sum_i \frac{-1}{2\Delta^2}\left(c_i^\dagger c_{i+1} + c_{i+1}^\dagger c_i \right) + \sum_i \left[ V_1 \cos\left( \frac{2\pi x_i}{a_1} \right) + V_2\cos\left( \frac{2\pi x_i}{a_2}\right) +\frac{1}{\Delta^2} \right]n_i + \frac{1}{2}\sum_{i,j}U(x_i-x_j) n_in_j.
	\end{split}
\end{equation}
In our paper, we choose $L=63~a_1$, $\Delta =L/1000$, $V_1=-3.5~E_h$ and the soft cutoff $d/a_1=0.05$. The choice of $V_2$ is important to obtain the flatbands, specifically the bandwidth should be much smaller than the band gap. As from Fig.~\ref{fig:chooseV2} showing the bandwidth and bandgap of a translational-invariant moir\'{e} lattice with $M=7$, the lowest band can be considered flat for $V_2/V_1 \gtrapprox 0.3$. Therefore, we fix $V_2=V_1=-3.5~E_h$ for our numerical simulation. To compute the ground state, we minimize the energy of the discretized model using the density matrix renormalization group \cite{Stoudenmire2012} implemented by the iTensor library \cite{itensor}. 

\begin{figure}[h!]
	\begin{minipage}{0.45\textwidth}
		\includegraphics[width=\textwidth]{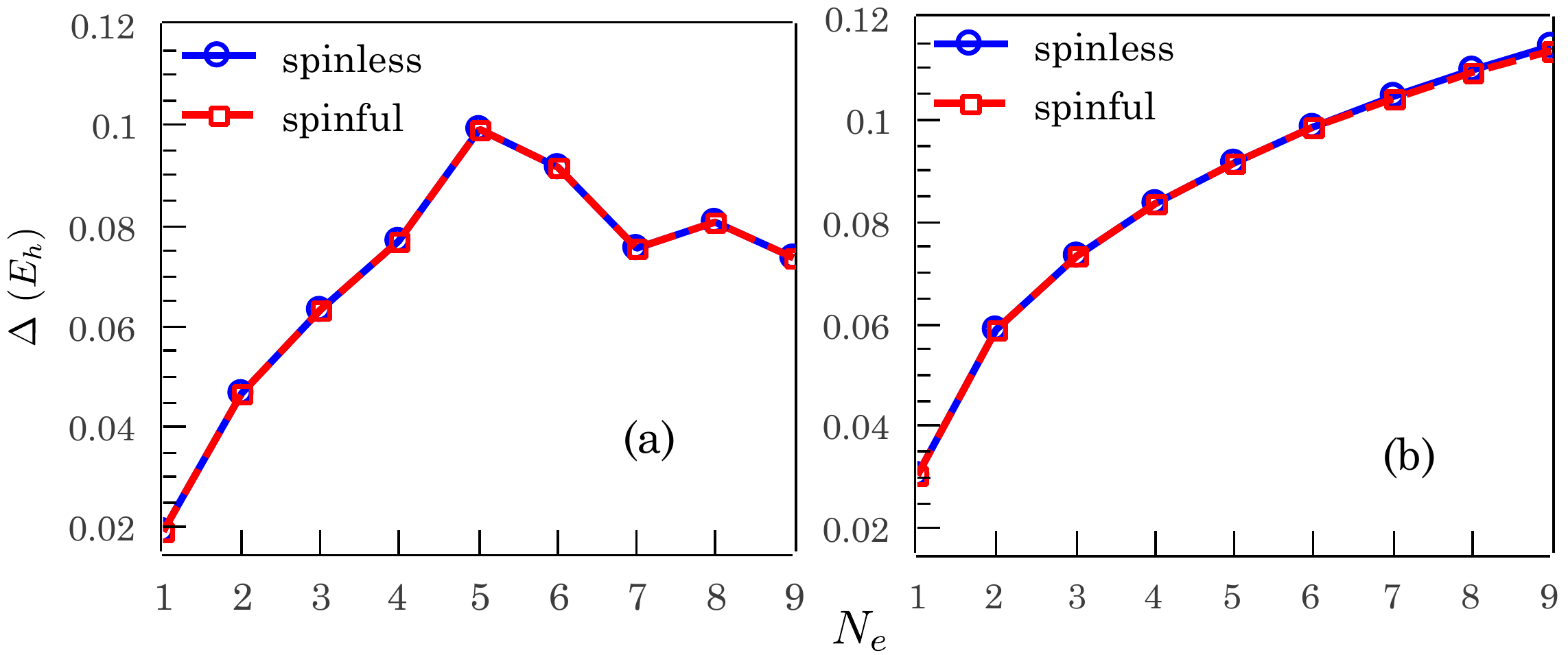}
	\end{minipage}
	
	\caption{Charge gap calculated from interacting spinless and spinful tight-binding models. (a) Superlattice $M=7$ with 10 sites, $U_c = 1/7~E_h$, $t=1.2\times10^{-3}~E_h$. (b) Original lattice $M=1$ with 70 sites, $U_c=1~E_h$, $t=0.4~E_h$.  In both cases, the exchange energy difference between the spinful and spinless models is negligible.\label{fig8}}
\end{figure}

To test various aspect of the continuum model ground state, we utilize a simplified but equivalent tight-binding model
\begin{equation}
	H_{tb}=\sum_{i} t(c_i^\dagger c_{i+1}+c_{i+1}^\dagger c_i) + 2U_c n_{i\uparrow}n_{i\downarrow} +\frac{1}{2}\sum_{i,j} \frac{U_c}{|j-i|} n_in_j,
\end{equation} 
where $U_c$ and $t$ are obtained from solving the moir\'{e} lattice in $k-$space. First, we can use the tight-binding model to justify the use of the spinless continuum model. In the insulating phase, the exchange energy is exponentially small with spin playing no role on the physics of gaps and density distributions of interest. From Fig.~\ref{fig8} where we compute the charge gaps for the spinless and spinful cases, it is clear that our results presented in this work are unaffected by the spinless approximation. 

\begin{figure}[h!]
	\centering
	\begin{minipage}{0.02\textwidth}
		\rotatebox{90}{$\Delta E$}
	\end{minipage}
	\begin{minipage}{0.40\textwidth}
		\centering
		\includegraphics[width=\textwidth]{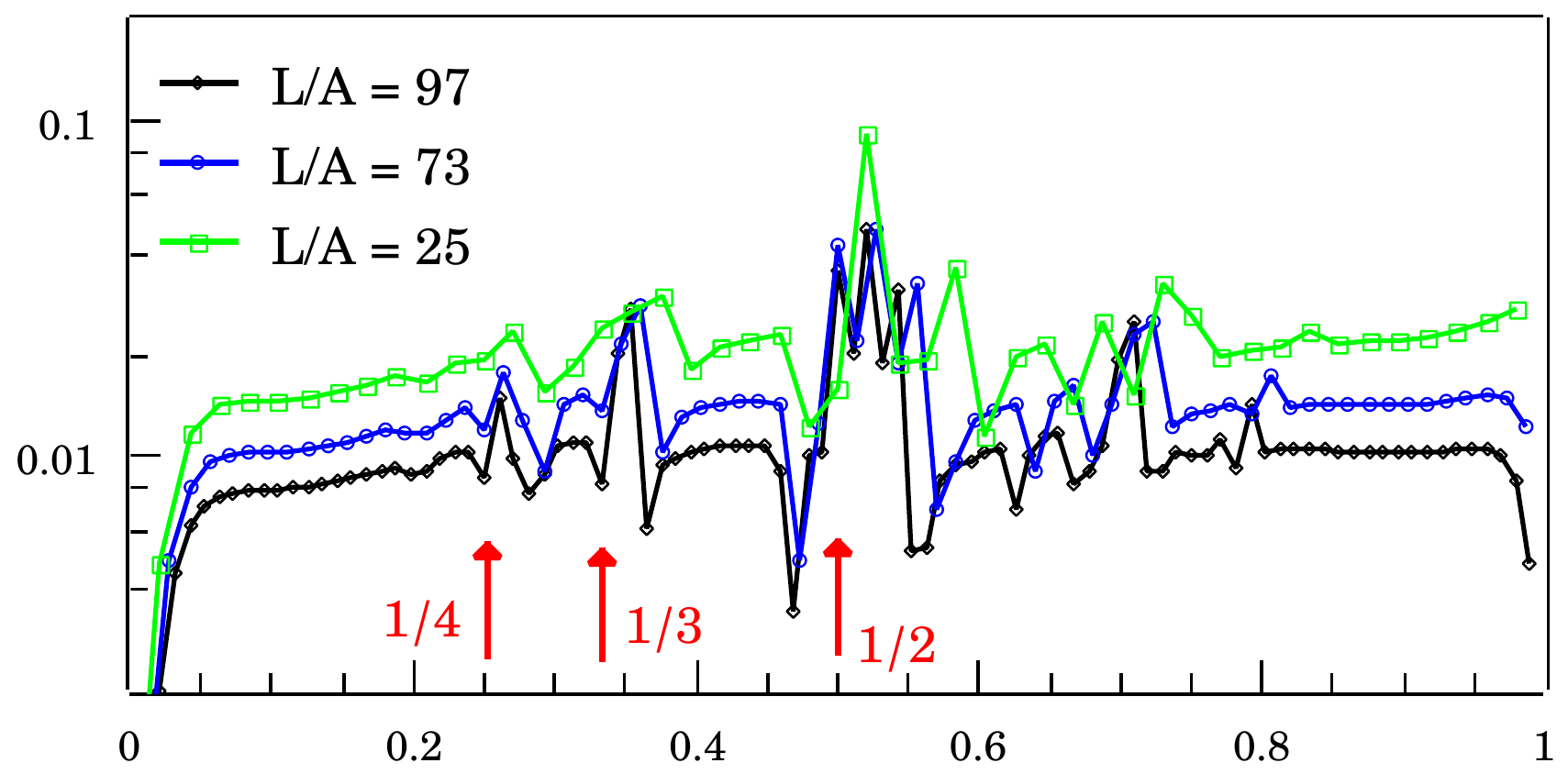}
		
		Filling factor $\nu$  	 		
	\end{minipage}
	\caption{Charge gap of moir\'{e} pattern $M=7$ at different system sizes.\label{fig:checkgap}}
\end{figure}

Next, we can check the existence of other charge gaps at $1/n$ fillings for larger system sizes by the tight-binding model. In Fig.~\ref{fig:checkgap}, we show that a series of prominent charge gaps pined to the $1/n$-fillings can be seen at different system sizes. For small systems, each added electron changes the filling factor significantly, thus decreasing the visibility of $1/n$-gaps with high $n$ (low filling).

\section{Moir\'{e} length in finite size systems}
\begin{figure}[h!]
	\centering
	\begin{minipage}{0.02\textwidth}
		\rotatebox{90}{\hspace{0.2in} $\braket{U}/r_s$}
	\end{minipage}
	\begin{minipage}{0.42\textwidth}
		\centering
		Approximated $M^*$
		\includegraphics[width=\textwidth]{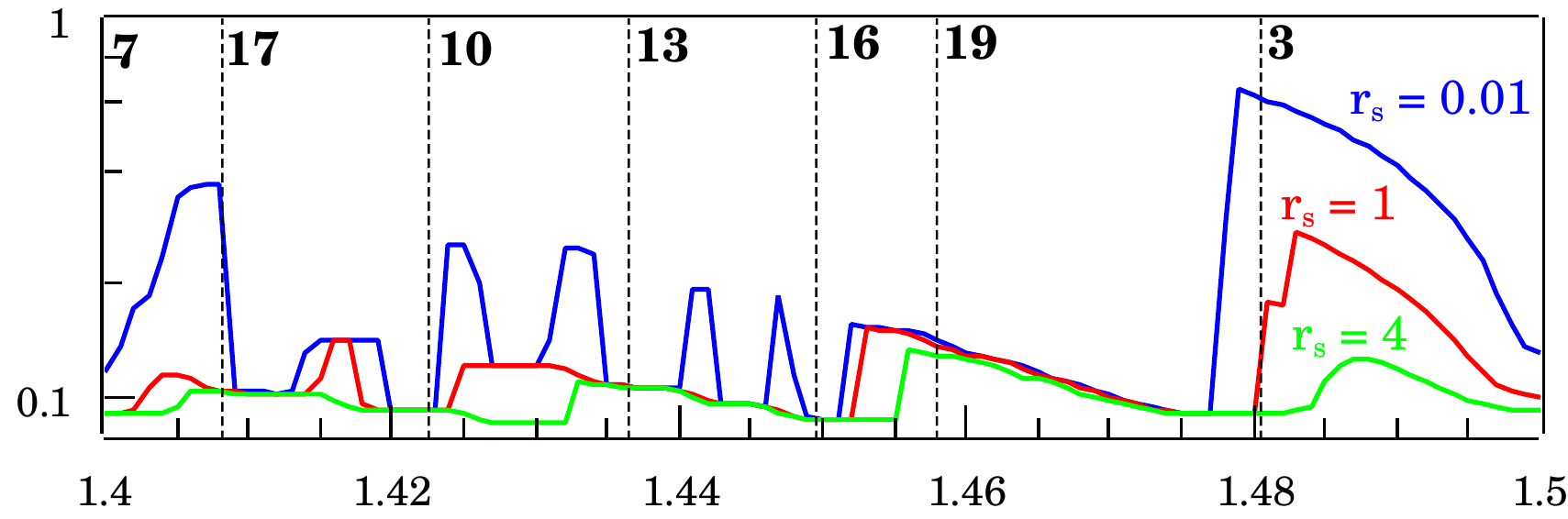}
		$m$ 	 		
	\end{minipage}
	\caption{Average interaction energy normalized against $r_s$ where the peaks indicate the suppression of the Mott phase and usually coincide with the domain boundary between approximated values of $M^*$.  \label{fig:boundary}}
\end{figure}

If the moir\'{e} ratio $m=a_2/a_1$ is an irrational number or a rational number $M/P$ but $Ma_1>L/N$, we cannot construct an effective periodic moir\'{e} pattern. In that case, we approximate $m$ by rational number $M^*/P^*$ so that $M^*a_1<L/N$ and the residue $m-M^*/P^*$ contributes as a quasi-disordered potential. The detailed process involves minimizing $\left|m-M^*/[M^*/m] \right|$ for $M^*$ ranging from 1 to $\lfloor L/(Na_1) \rfloor$ where $[x]$ and $\lfloor x \rfloor$ are the nearest integer to $x$ and the largest integer less than $x$. In Fig.~\ref{fig:boundary}, we calculate the scaled average interaction energy with the boundary between domains of approximated $M^*$. The disordered effect is strongest at the superlattice size domain boundary, which can replace the Mott configuration and increase the interaction energy significantly, especially for weakly interacting cases. 
\end{document}